\documentclass[pra,aps,superscriptaddress,longbibliography]{revtex4} 
\usepackage{graphicx,color,appendix}
\usepackage{setspace,appendix} 
\usepackage{amsmath}
\usepackage{amssymb,braket}
\usepackage{ulem}
\usepackage[colorlinks = true]{hyperref}

\newcommand{\bla}{\color{black}}

\usepackage{titlesec}
\usepackage[T1]{fontenc}

\begin{document}
	\title{A class of quasi-eternal non-Markovian Pauli channels and their measure}
	\author{Shrikant Utagi}
	\email{shrik@poornaprajna.org}
	\affiliation{Theoretical Sciences Division, Poornaprajna Institute of Scientific Research, Bidalur, 	
		Bengaluru- 562164, India}
	\affiliation{Graduate Studies, Manipal Academy of Higher Education, Manipal -576104, India.}
	\author{Vinod N Rao}
	\affiliation{Theoretical Sciences Division, Poornaprajna Institute of Scientific Research, Bidalur, 	
		Bengaluru- 562164, India}
	\author{R. Srikanth}
	\email{srik@poornaprajna.org}
	\affiliation{Theoretical Sciences Division, Poornaprajna Institute of Scientific Research, Bidalur, 	
		Bengaluru- 562164, India}
	\author{Subhashish Banerjee.}
	\email{subhashish@iitj.ac.in}
	\affiliation{Indian Institute of Technology, Jodhpur- 342037, India.}	

\begin{abstract}
We study a class of qubit non-Markovian general Pauli dynamical maps with multiple singularities in the generator. We discuss a few easy examples involving trigonometric or other non-monotonic time dependence of the map, and discuss in detail the structure of channels which don't have any trigonometric functional dependence. We demystify the concept of a singularity here, showing that it corresponds to a point where the dynamics can be regular but the map is momentarily non-invertible, and this gives a basic guideline to construct such non-invertible non-Markovian channels. Most members of the channels in the considered family are quasi-eternally non-Markovian (QENM), which is a broader class of non-Markovian channels than the eternal non-Markovian channels. In specific, the  measure of quasi-eternal non-Markovian (QENM) channels in the considered class is shown to be $\frac{2}{3}$ in the isotropic case, and  about 0.96 in the anisotropic case.
\keywords{Quantum depolarizing channels \and Quasi-eternal non-Markovianity \and Singularities}
\end{abstract}

\maketitle

\section{Introduction \label{intro}}

It is well known that noise is an inevitable attendant feature of quantum systems in a realistic scenario \cite{reich2015exploiting,liu2011experimental,chruscinski2010non}. Quantum non-Markovianity is a memory property that can help the system  recover coherence through environment-induced decoherence \cite{breuer2002theory,banerjee2018open}. In that sense, it can be thought of as a resource for an efficient real-life implementation of quantum information processing tasks \cite{bylicka2014non-Markovianity}, such as quantum key distribution \cite{shrikant2020pingpong,thapliyal2017quantum}, quantum teleportation of mixed quantum states \cite{laine2014non-local}, and quantum thermodynamics \cite{george2018thermodynamics}, to name a few.  Therefore, the characterization and study of non-Markovian quantum channels continues to be a major research activity \cite{li2019non-Markovian,naikoo2019facets,naikoo2019coherence,naikoo2019quantumness,shrikant2020temporal}. 

In many of the works in the literature, the study of non-Markovian dynamics has been done by incorporating physical models of particular interest \cite{grabert1988quantum,sbsterngerlach,sbqbm,kumar2017nonmarkovian}. Here, we provide a model-independent framework, introduced in \cite{shrikant2018non-Markovian}, in which a method of creating non-Markovian analogues of dephasing and depolarizing channels was presented. We adopt that method to obtain a family of general Pauli dynamical maps, which it will be convenient to split into the two classes-- of the isotropic and anisotropic non-Markovian depolarizing channels, which are the special instances of generalized Pauli channels, which in turn are special cases of random unitary channels.  The method is tailored so that the non-Markovianity is associated with the presence of singularities in the time-local generator. Often it will be convenient to simplify the description of the decoherence using a noise parameter $p(t)$ instead of time $t$ itself \cite{shrikant2018non-Markovian}, whose functional form is not needed for the purpose of this paper. For example, in the dephasing channel given by $\mathcal{E}_t[\rho(t)] = (1-p(t))\rho(0)+ p(t) \sigma_z \rho(0) \sigma_z$, $t = \{0,\infty\}$ corresponds to $p = \{0,\frac{1}{2}\}$ . Throughout this paper, we associate time-dependence of states, operators and functions with $p$. 

The appearance of singularities in the time-local generator is not new. Phenomenological models such as a two-level system interacting with a bosonic reservoir \cite{breuer2002theory} constitute an example of an instance in which the time-local generator in the canonical master equation has infinite singularities.

This paper is arranged as follows. In Section \ref{sec:mNmconditions} we review non-Markovianity conditions for  generalized Pauli channels and dwell briefly on the existence of singularities. In Section \ref{sec:examples}, we give a few easy examples of channels that have an infinite number of singularities, in order to motivate the non-trivial character of the channels introduced in the next section. In Section \ref{sec:asymDepol}, we present depolarizing channel with anisotropic non-Markovianity as an example of a generalized Pauli channel, which in turn is a special case of random unitary dynamics. An analysis of the singularities in the master equation of the channel is presented in Section \ref{sec:singul}, along with a demonstration of the fact that these singularities are non-pathological.  In Section \ref{sec:nmDepol} we revisit the isotropic version of the non-Markovian depolarization channel, and quantify its non-Markovianity using a measure due to Hall-Cresser-Li-Anderson (HCLA) \cite{hall2014canonical}, as well as a recently introduced \cite{shrikant2020temporal} measure. In  Section \ref{sec:enm} we  show that most of the channels studied here are {\it quasi-eternally non-Markovian} (QENM), i.e., they have negative decoherence rate after a finite time $t^-$ (which corresponds to a singularity), but may be completely positive (CP)-divisible before that \cite{santis2019witnessing}. Then we conclude in Section \ref{sec:conclu}.
A list of notations used is summarized in the Table in Appendix \ref{sec:appendixB}. 

\section{Non-Markovianity conditions for Pauli dynamical maps \label{sec:mNmconditions}} 
A generalized Pauli channel \cite{wudarski2013non-Markovian} is given by the operator-sum (or, Kraus) representation $\mathcal{E}(t)[\rho] = \sum_j K_j(t) \rho K_j(t) ^\dagger $ with $K_j(t) \equiv \sqrt{\kappa_j(t)}\sigma_j$ being the Kraus operators, where $\sigma_0=I$ and $\sigma_1,\sigma_2,\sigma_3$ are Pauli $ X, Y $ and $ Z $ operators, respectively. By projecting the map onto the Pauli matrices, we obtain the operator eigenvalue equation $\mathcal{E}(t)[\sigma_i] = \nu_i(t)\sigma_i$, where $\nu_i(t)$ are the time-dependent eigenfunctions of the Pauli map given by 
\begin{align}
\nu_j(t) = \sum_{i=0}^{3} H_{ji} \kappa_i(t),
\label{eq:lambdas}
\end{align} and
\begin{align}
H =	\left(
\begin{array}{cccc}
1 & 1 & 1 & 1 \\
1 & 1 & -1 & -1 \\
1 & -1 & 1 & -1 \\
1 & -1 & -1 & 1 \\
\end{array}
\right)
\end{align} is the Hadamard matrix. The decay rates are generally obtained as \cite{wudarski2013non-Markovian}
\begin{align}
\gamma_i(t) = \frac{1}{4}\sum_{j=0}^{3} H_{ij} \frac{\dot{\nu}_j(t)}{\nu_j(t)}.
\label{eq:rates0}
\end{align} 
Assuming that intermediate maps $\mathcal{E}(t_2, t_1)$ for all $t_2 < t_1 < t_0 $, are invertible, the canonical form of master equation corresponding to the map (\ref{eq:paulimap}) is 
\begin{align}
\dot{\rho} =& \mathcal{L}(t) [\rho] \nonumber \\ =& \sum_{j=1}^{3} \gamma_j(t) (\sigma_j \rho \sigma_j^\dagger - \rho), 
\label{eq:masterpauli}
\end{align} 
where $\gamma_j (t)$ is the decoherence rate corresponding to the $j^{\rm th} $ unitary operation in the channel.    $\mathcal{L}(t) $ is the generator of the dynamics $\mathcal{E}$ such that $ \mathcal{L}(t) = \mathcal{\dot{E}}(t) \mathcal{E}(t)^{-1}$. 

Assuming that the channel is invertible, the necessary condition for it to be completely positive CP-divisible \cite{RHP10} is:
\begin{align}
\gamma_1(t) \ge 0, \quad  \gamma_2(t) \ge 0, \quad \gamma_3(t) \ge 0,
\label{eq:CP-div}
\end{align}
i.e., the positivity of Lindblad rates that appear in the time-local Gorini-Kossakowski-Lindblad-Sudarshan (GKSL) form such as Eq. (\ref{eq:masterpauli}). 
A necessary condition for positive P-divisibility (or equivalently of Markovianity according to the criterion due to Breuer-Laine-Piilo (BLP) \cite{breuer2009measure}, in the qubit case) is \cite{wudarski2013non-Markovian}
\begin{align}
\gamma_1(t) + \gamma_2(t) \ge 0, \nonumber \\
\gamma_2(t) + \gamma_3(t) \ge 0, \nonumber \\
\gamma_3(t) + \gamma_1(t) \ge 0.
\label{eq:P-div}
\end{align} 
However, if the channel is non-invertible, as in the present case, then the condition Eq. (\ref{eq:CP-div}) is only sufficient but not necessary for CP-divisibility \cite{chruscinski2018divisibility}. Specifically, the rate may be temporarily negative after a singularity even though the channel is CP-divisible. 

Note that at the point of singularity in the time-local generator, the rates in the GKSL equation become infinite and the map is momentarily non-invertible. In the present case, these singularities are finite in number and non-pathological \cite{shrikant2018non-Markovian}, and the above criteria can be applied at all points except these.

From Eqs. (\ref{eq:CP-div}) and (\ref{eq:P-div}), we find that P-indivisibility implies CP-indivisibility, but not vice versa. The conditions under which these two criteria are equivalent is investigated in Ref. \cite{chruscinski2018divisibility}.

From Eq. (\ref{eq:rates0}) it follows that a generator possesses a singularity whenever $\nu(t)=0$. Later we will show that if $\nu(t^\prime)=0$ in finite time $t^\prime$, then the decay rate flips sign, thereby giving rise to non-Markovian evolution following the singularity at the time $t^-$. 

Significantly, we shall point out that the singularities are non-pathological \cite{shrikant2018non-Markovian, jagadish2019measure}, in that the dynamics can be regular at the instant of singularity $t^\prime$. In fact, the existence of such a singularity in the generator $\mathcal{L}(t)$ corresponds to a momentary indistinguishability of the evolved version of distinct states in the sense of BLP \cite{breuer2009measure}, and thereby the momentary non-invertibility of the dynamical map $\mathcal{E}(t)$ in the sense of CP-divisibility according to the Rivas-Huelga-Plenio (RHP) criterion \cite{RHP10}. For purposes of quantification, the singularities may be suitably normalized.

For any qubit Pauli map with multiple decoherence rates, it was shown \cite{sagnik2019information} that non-Markovianity criterion based on distinguishability and CP-divisibility are equivalent. Therefore, we associate the non-Markovianity of channels in this work with the negativity of one or more decay rates in the canonical master equation. 

\section{Elementary examples of non-Markovian Pauli channels with singularities \label{sec:examples}}
In this section we show how one can straightforwardly obtain an infinite number of singularities in the generator with oscillatory channel decay probabilities. As a quick example, consider a dephasing channel with Kraus operators: 
\begin{align}
K_0(t) = \sqrt{\frac{1+p(t)}{2}} \sigma_0 \quad ; \quad K_1(t) = \sqrt{\frac{1-p(t)}{2}} \sigma_3,
\label{eq:template}
\end{align} where $p(t) = \cos(\omega t)$. Here, $\sigma_0=I$ is the identity operator and $\sigma_3$ is Pauli Z operator. A singularity corresponds to the point of maximal dephasing, which collapses all states on the same azimuth of the Bloch sphere.  The decay rate of the channel reads $\gamma(t) = - \frac{1}{2 p(t)}\frac{d p(t)}{dt}=  \frac{\omega}{2}\tan (\omega t)$, which has infinite singularities, corresponding to time being odd multiples of $\frac{\pi}{2\omega}$. As a second easy example, consider a Pauli channel with the Kraus operators:
\begin{align}
K_0 &= \sqrt{\frac{1+ 3\cos (\omega t)}{4}} \sigma_0, \quad
&K_1 = \sqrt{\frac{1-\cos(\omega t)}{4}}\sigma_1,\nonumber\\
K_2 &= \sqrt{\frac{1-\cos (\omega t)}{4}} \sigma_2, \quad
&K_3= \sqrt{\frac{1-\cos (\omega t)}{4}} \sigma_3.
\end{align} 
Singularities correspond, as above, to times where $\cos(\omega t)$ vanishes, leading to maximal mixing, and hence momentary irreversibility. The above maps have time-dependent eigenvalues $\nu_1(t)=\nu_2(t)=\nu_3(t)=\cos(\omega t)$, and the decay rates (\ref{eq:rates0}) are found to be $\gamma_1(t)=\gamma_2(t)=\gamma_3(t)= \frac{\omega }{4}  \tan ( \omega t  )$. 

The concept of eternal non-Markovianity was introduced in \cite{hall2014canonical}, where a Pauli channel with decay rates $\gamma_1(t)=\gamma_2(t)=1$ and $\gamma_3(t)=-\tanh(t)$ was proposed as an example of a CP-indivisible channel that is Markovian according to the distinguishability criterion \cite{breuer2009measure}. As evident, this channel has the property that it has a negative rate for all times $t>0$, and accordingly called ``eternally non-Markovian''.

Following the definition proposed in \cite{santis2019witnessing}, we shall refer to a channel $\mathcal{E}$ as ``quasi-eternally non-Markovian'' (QENM) if there exists a finite time $t^-$ such that $\mathcal{E}$ is CP-indivisible for all times $t > t^-$, i.e., if there exists a finite $p^-$ such that the channel is CP-indivisible for all $p \in [p^-, \frac{3}{4}]$. The channel may be CP-divisible for $t \le t^-$.  We remark that this differs from the notion of quasi-eternal non-Markovianity proposed in Ref. \cite{rivas2017weak}, where this term refers to a dynamics that is CP-indivisible for time $t<t^\circ$, and $t^\circ$ is the instant at which the system has almost reached its steady state.\bla

In Eq. (\ref{eq:template}), setting $p(t) = 1-t e^{1-t}$, we find that this function gives an easy example of a dephasing channel that is quasi-eternally non-Markovian (QENM), with the rate $\gamma(t) =  -\frac{\dot{p}}{2 p} = \frac{e (t-1)}{e t-e^t}$, attaining a singularity at the point $t$ where $et = e^t$. The rate is positive before the singularity and negative after it. 

The above examples are manifestly maps that are non-monotonic, in that the mixing functions such as $p(t)$ are non-monotonic. This is in consonance with the usual notion of associating non-Markovian behavior with break in monotonicity. However,  in this work, we will discuss a class of completely positive trace preserving (CPTP) Pauli dynamical maps that are monotonic. Specifically, this entails that time-dependent eigenvalues  $\nu(t)$ are monotonic functions. The origin of non-Markovianity here is that the maps evolve the state beyond the point of maximum decoherence, leading to negative decay rates. The maps so produced are naturally quasi-eternal non-Markovian (QENM) in the sense mentioned above.


\bla	\section{Anisotropic non-Markovian depolarization \label{sec:asymDepol}}	
In the following, we adopt the framework of \cite{shrikant2018non-Markovian}, where non-Markovian analogues of Pauli dephasing and depolarizing channels were introduced. Here, in terms of the parameter $p$, the map of a qubit Pauli channel is given by:
\begin{align}
\mathcal{E}(p) [\rho] = \sum_{j=0}^{3} \kappa_j (p) \sigma_j \rho \sigma^{\dagger}_j ,
\label{eq:paulimap}
\end{align}
where $ \sigma_0 = I $, and $\sigma_j, \; j \in \{1,2,3\}$ are Pauli operators, and $ \sum_{j=0}^{3} \kappa_j (p) = 1$.  Henceforth, it will be convenient to use the notation where a monotonic parameter $p(t)$ is used in place of time $t$. This is done essentially because the detailed functional form of $p(t)$ does not affect the results here. 

Using the method introduced in \cite{shrikant2018non-Markovian}, one may generate non-Markovian extensions
of Kraus operators $K_j$'s as:
\begin{align}
K_0 &= \sqrt{(1+\Lambda_0)(1-p)} \sigma_0, \quad
K_1 = \sqrt{(1 + \Lambda_1) \frac{p}{3}}\sigma_1,\nonumber\\
K_2 &= \sqrt{(1 + \Lambda_2) \frac{p}{3}} \sigma_2, \quad
K_3= \sqrt{(1 + \Lambda_3) \frac{p}{3}} \sigma_3,
\label{eq:nmdepol}
\end{align}
where $\Lambda_j$ ($j  \in \{0,3\}$) is a real function,  and $p$ is a
real parameter, which acts like time, in this framework. It rises monotonically from  0 to $\frac{3}{4}$, and its functional form does not matter here. 
The variables $\Lambda_j$ satisfy the following condition 
\begin{align}  (1-p) \Lambda_0 \, +  \frac{p}{3} (\Lambda_1 + \Lambda_2 +\Lambda_3)
\, = 0,
\label{eq:complete}
\end{align}
as a consequence of the completeness requirement $ \sum_j K^{\dagger}_jK_j  = I$. Moreover, the channel is ensured to be completely positive in the sense that the Choi matrix $\chi = (\mathcal{E}(p_2,p_0) \otimes I)[\rho^{+}]$ is positive semidefinite i.e., $\chi \ge 0$ for all $p_2 > p_1 > p_0$, where $\rho^{+}= (\ket{00}+\ket{11})(\bra{00}+\bra{11})$ is a maximally entangled state. And the channel is said to be CP-\textit{divisible} if the Choi matrix $\chi$ of the intermediate map $\mathcal{E}(p_2,p_1)$ for all $p_2> p_1$ is positive, which will be discussed in Section (\ref{sec:choi}).  In  agreement with  Eq.   (\ref{eq:complete}), we  make the  following
choices: $ \Lambda_0 = - (l+m+n)  p$, $ \Lambda_1 = 3 l (1-p)
$, $ \Lambda_2 = 3 m(1-p)$ and $ \Lambda_3 = 3 n (1-p)$, where  $l,m,n$ are  real. Then,  the non-Markovian  Kraus operators take the form 
\begin{align}
K_0(p)  &= \sqrt{[1  - (l+m+n)  p](1-p)} \;  \sigma_0,\nonumber  \\ 
K_1(p)  &= \sqrt{[1 + 3l (1-p)]\frac{p}{3}} \; \sigma_1, \nonumber\\
K_2(p)  &= \sqrt{[1 + 3m (1-p)]\frac{p}{3}} \; \sigma_2, \nonumber\\
K_3(p)  &= \sqrt{[1 + 3n (1-p)]\frac{p}{3}} \; \sigma_3. 
\label{eq:nmdepol2}
\end{align} 
The  parameters  $l,m,n$  may   be  seen  to   represent  the
non-Markovian behavior of the channel, such that setting $l =m=n= 0$
reduces the  Kraus operators in  Eq.  (\ref{eq:nmdepol2}) to  those in
the conventional Markovian depolarizing channel, which is a type of Pauli channel. It is easy to see that setting $l=m=n \equiv \alpha$ reduces (\ref{eq:nmdepol2}) to the (isotropic) non-Markovian depolarizing channel introduced in \cite{shrikant2018non-Markovian}. 

By definition, $\mathcal{E}(p=0)=\mathbb{I}$, which implies that $K_0(0) = 1 $. 
In a depolarizing channel, the mixing parameter $p(t)$ varies from $0$ to $\frac{3}{4}$, at which point the state becomes maximally mixed. For $p >\frac{3}{4}$, the system deviates from $\frac{\mathbb{I}}{2}$, i.e., to re-cohere. 

The canonical decay rates $ \gamma_j(p)$ in the equation (\ref{eq:masterpauli}) are calculated using the method given in \cite[Section 2]{wudarski2013non-Markovian}, and are found to be:
\begin{align}
\gamma_1(p) &= \frac{1}{4} \left(- \frac{\dot{\nu}_1}{\nu_1}  - \frac{\dot{\nu}_2}{\nu_2}+\frac{\dot{\nu}_3}{\nu_3}\right), \nonumber \\ 
\gamma_2(p) &= \frac{1}{4} \left(- \frac{\dot{\nu}_1}{\nu_1}  + \frac{\dot{\nu}_2}{\nu_2} -\frac{\dot{\nu}_3}{\nu_3} \right), \nonumber \\ 
\gamma_3(p) &= \frac{1}{4} \left(\frac{\dot{\nu}_1}{\nu_1}  - \frac{\dot{\nu}_2}{\nu_2} - \frac{\dot{\nu}_3}{\nu_3}\right),
\label{eq:asymgammas}
\end{align} 
where 
\begin{subequations}
\begin{align}
 \dot{\nu}_j &= 
6x_j(2p -1) - 4, \label{eq:num} \\
\nu_j &= 
6px_j(p - 1) - 4p+3. \label{eq:den}
\end{align}
\label{eq:denominators}
\end{subequations}
 where $\dot{\nu} = \frac{d\nu}{dp}$. 
Here $x_1 \equiv l+m, x_2 \equiv l+n$ and $x_3 \equiv m+n$.
\begin{center}
\begin{figure}[hb!]
	\centering
	\includegraphics[width=7cm]{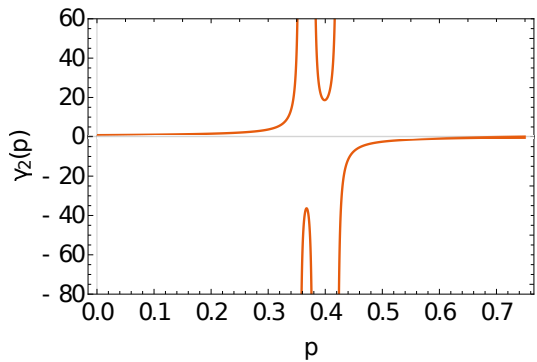}
	\caption{Plot of the decay rate $\gamma_2(p)$, in Eq. (\ref{eq:asymgammas}), for the values of $l=0.4$, \,$m=0.5$ and $n=0.65$. The singularities occur at $ p^-_2 \approx 0.356$, $ p^-_3 =0.378 $ and $p^-_1 = 0.42$, respectively.}
	\label{fig:gamma2}
\end{figure}	
\end{center}

\section{Analysis of singularities \label{sec:singul}} 

We now consider the questions of how the non-Markovian parameters $l,m$ and $n$ determine the location of the singularities and thereby the nature of the intermediate map, and finally the physical interpretation of the singularities. 

\subsection{Location of singularities}
In order to analyze the singularities, one needs to extract the roots of  Eq. (\ref{eq:denominators}). The zeros of $\nu_j$ in Eq. (\ref{eq:asymgammas}) yield the singularities of the decay rates. Solving for any $\nu_j$, one finds:
\begin{align} 
p^{\pm}_j = \frac{2 + 3 x_j \pm \sqrt{9 x_j^2 - 6 x_j + 4}}{6 x_j}, 
\label{eq:singuasym}
\end{align} 
$p_j^\pm$ are the two solutions for $p$ in the second equation of Eq. (\ref{eq:denominators}) for a given $j$.
The larger of the two roots can be ignored as it appears after $p=\frac{3}{4}$.
Here, $l=m=n=0$ corresponds to Markovian evolution, and if any one of the parameter is non-zero, the channel exhibits non-Markovian evolution after it experiences a singularity. 

\begin{figure}
	\centering
	\includegraphics[width=6.5cm]{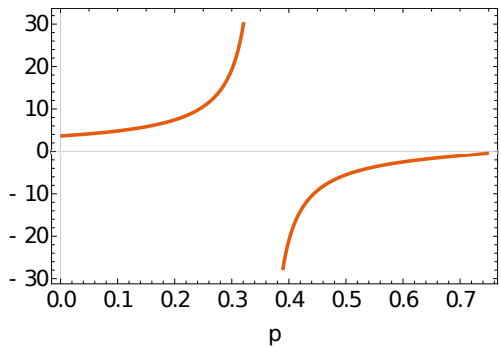}
	\caption{Plot of  $-\frac{\dot{\nu}_j(x)}{\nu_j(x)}$  for the value of $x_j=m+n$, for the values of $m=0.5$ and $n=0.65$ as was the case considered in Fig.  (\ref{fig:gamma2}). The singularity  in Eq. (\ref{eq:singuasym}), occurs at $p^- \approx 0.356$, which is the first singularity to appear in Fig. (\ref{fig:gamma2}).}
	\label{fig:ratio_singu}
\end{figure}

The singularity pattern in Figure \ref{fig:gamma2} can be understood by following the behavior of any function
$\frac{\dot{\nu}_j}{\nu_j} = \frac{\dot{\nu}_j}{(p^-_j-p)(p^+_j-p)}$ in Eqs. (\ref{eq:asymgammas}). This immediately shows that the function flips its sign at $p=p_j^-$. An example is plotted in Figure (\ref{fig:ratio_singu}). 

Referring to the form of Eq. (\ref{eq:asymgammas}), this implies that each rate by itself will have its sign flipped at the singularities. To see this, assuming more than one singularity, note that close to one of the singularities, only the function(s) $\frac{\dot{\nu}_j}{\nu_j}$ that diverges at that singularity will be unbounded, whereas the other contribution(s) to the given rate $\gamma_k$ will be finite. Thus, the rate as a whole inherits the sign flipping behavior. Figure \ref{fig:gamma2} exemplifies this pattern. Here the behavior of the decay rate of $\gamma_2(p)$ is the composite of the above pattern of evolution of the three terms $\frac{\dot{\nu}_j}{\nu_j}$. 

If $l \ne m \ne m$, then $x_1 \ne x_2 \ne x_3$, leading to three singularities, whilst if two of the non-Markovianity parameters are equal, say $l=m \ne n$, with $l, m, n > 0$, then there are two singularities at $p^{-}_1 = p^-_3$ and at $p^{-}_2$. If $l=m=n > 0$, then a single singularity occurs at $p^{-}_1= p^{-}_2 = p^{-}_3=0$.

\begin{figure}
	\centering
	\includegraphics[width=6.5cm]{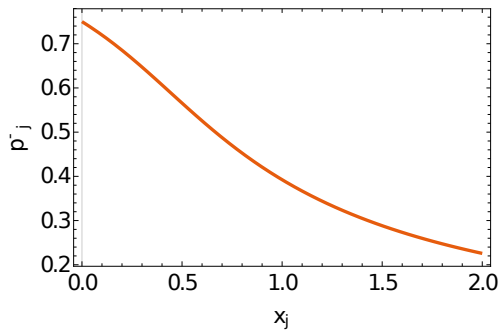}
	\caption{Plot of the position of the singularity $p_j^-$ Eq. (\ref{eq:singuasym}) against $x_j$, where $x_1 \equiv l+m, x_2 \equiv l+n$ and $x_3\equiv m+n$. Note that each $x_j$ corresponds to a sheet in the $lmn$ parameter space.}
	\label{fig:pstar}
\end{figure}

The function $p^-_j(x)$ monotonically falls with $x$, which varies in the interval $[0,2]$ (Figure \ref{fig:pstar}). In particular
$$ \lim_{x \to 2} p^-_j(x) = \frac{2}{3} - \frac{\sqrt{7}}{6}  \le p \le \frac{3}{4} = \lim_{x \to 0} p(x),$$ 
showing that as the non-Markovianity parameters $l, m$ or $n$ are increased, the position of any singularity here shifts monotonically to the left.

\subsection{(Not) complete positivity of the intermediate map}

\begin{figure}
	\centering
	\includegraphics[width=7cm]{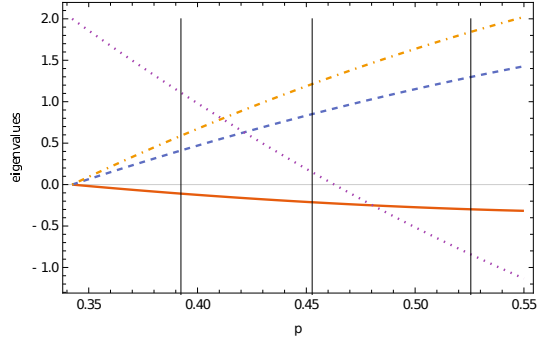}
	\caption{(Color online) Plot of the eigenvalues of the intermediate map, Eq. (\ref{eq:multi}), for the time interval $[s,p]$, against the values of $p \in (s, \frac{3}{4}]$, with $l:=0.2$, $m:=0.4$, $n:=0.6$. Here, $s$ is taken to be slightly smaller than  $p^-_{\rm min} \equiv \textrm{min}\{p^{-}_1,p^{-}_2,p^{-}_3\}$. For the above $l, m$ and $n$ values, $p^{-}_{\rm min} = p^-_3= 0.392$. The vertical lines show the positions of the singularities (the second and third singularities appearing at $p^{-}_2 = 0.452$ and $p^{-}_1 = 0.525$, respectively). We note that $\lambda_1$ (bold, red curve) becomes negative in the approach to $P^-_{\rm min}$, signaling the onset of CP-indivisibility just before to first singularity. Correspondingly, the decay $\gamma_1(p)$ can be shown to become negative in this region.}
	\label{fig:multi2}
\end{figure}

In light of Ref. \cite{chruscinski2018divisibility}, negativity of a decoherence rate in a time interval following the first singularity will not necessarily imply CP-indivisibility. Therefore, to establish that the map is CP-indivisible in this region, we require to explicitly consider an intermediate map that should be not completely positive (NCP) and, in particular, that at least one of the eigenvalues of the Choi matrix of the intermediate map is negative. A specific instance of this is given in Figure (\ref{fig:multi2}), as discussed below.

A completely positive (CP) trace preserving map taking the quantum state from time $p_0$ to $p_2$ may generally be written as $\mathcal{E}(p_2, p_0) = \mathcal{E}(p_2,p_1)\mathcal{E}(p_1,p_0)$, where the intermediate map $\mathcal{E}(p_2,p_1)$ need not be CP. Assuming that the intermediate map is invertible, we have $\mathcal{E}(p_2,p_1) = \mathcal{E}(p_2,p_0)\mathcal{E}(p_1,p_0)^{-1}$. The time-dependent eigenvalues $\lambda_j(p_2,p_1)$ of the Choi matrix $\chi \equiv (\mathcal{E}(p_2,p_1)\otimes \mathbb{I})(\ket{00}+\ket{11})$ are given below: Let $[p^{\pm}_j;p,s] = \frac{(p^{-}_j - p)(p^{+}_j - p)}{(p^{-}_j - s)(p^{+}_j - s)}$, with $j \in \{1,2,3\}.$ Then 
\begin{align}
\lambda_1(p,s) & = \frac{1}{2}\left(1 - [p^{\pm}_1;p,s] - [p^{\pm}_2;p,s]+ [p^{\pm}_3;p,s] \right), \nonumber \\
\lambda_2(p,s) & = \frac{1}{2}\left(1 - [p^{\pm}_1;p,s] + [p^{\pm}_2;p,s]- [p^{\pm}_3;p,s] \right),\nonumber \\
\lambda_3(p,s) & = \frac{1}{2}\left(1 + [p^{\pm}_1;p,s] - [p^{\pm}_2;p,s]- [p^{\pm}_3;p,s] \right), \nonumber \\
\lambda_4(p,s) & = \frac{1}{2}\left(1 + [p^{\pm}_1;p,s] + [p^{\pm}_2;p,s]+ [p^{\pm}_3;p,s] \right),
\label{eq:multi}
\end{align}
where, $p_2 \equiv p$ and $p_1 \equiv s$, for all $p \ge s > 0$. Here, $p^{\pm}_j$, with $j = \{1,2,3\}$ are as given in Eq. (\ref{eq:singuasym}). The eigenvalues are plotted against time $p$ for a fixed $s$ in Figure (\ref{fig:multi2}). It can be shown that the decay rates obtained by Eq.(\ref{eq:rates0}) flip sign from negative to positive or vice versa at each singularity. An example of this behavior was discussed in the previous section, with respect to rate $\gamma_2(p)$, which flips sign from positive to negative at the singular point  $p^{-}_{\rm min} = 0.392$ (see in Fig. (\ref{fig:gamma2})).  Similarly, it can be shown that $\gamma_1(p)$ become increasingly negative as the singularity at $p^-_{\rm min}$ is approached, and flips sign to positive at the singularity, indicating the onset of CP-indivisibility prior to the singularity. This explains the negativity of the bold-red curve in the Fig. (\ref{fig:multi2}) before the singularity at $p^{-}_{\rm min}$.  This example brings out an interesting interplay between singularities, negativity of decay rates and CP-indivisibility in Pauli channels.

\subsection{Singularities are not pathological \label{sec:choi}}

An important point that was implicit earlier, and which we make explicit now, is that although the generator  in the GKSL equation has a singularity, yet the dynamics is regular.  
The singularity corresponds on the level of the map to a momentary indistinguishability of a family of initial states, and thus to the non-invertibility of the map, i.e., the momentary non-existence of $\mathcal{E}(p)^{-1}$, after which a recurrence occurs. Noting that generator is the dynamical map's logarithmic derivative, i.e., $ \mathcal{L}(p) = \mathcal{\dot{E}}(p) \mathcal{E}(p)^{-1}$, the singular nature of the generator may be understood as a consequence of the non-invertibility of the map, despite the latter's differentiability.

This can be verified easily in the case of the isotropic non-Markovian depolarizing channel mentioned above, by noting that the trace distance (TD) between two initial pair of states under our channels fall monotonically continuously for $\alpha$=0. On the other hand, for $\alpha >0$, in the case of the isotropic depolarizing channel,  at the point of singularity, given by $p=p^-$ (Eq. (\ref{eq:singudepol})), TD vanishes, after which they again increase. However, the time evolution of TD is continuous even at $p = p^-$, and in that sense, the singularity is not pathological.  

We illustrate this aspect with the following mathematical arguments.
The final state obtained after the channel acts on a generic state:
\begin{align}
&\mathcal{E}^{\rm \tiny iso}
\left(
\begin{array}{cc}
a & b \\
b^* & 1-a \\
\end{array}
\right) 
= \left(
\begin{array}{cc}
A  & B \\
B^* & 1-A\\ \label{eq:stateisotropic}
\end{array}
\right),
\end{align}
where 
\begin{align}
A &= a+\frac{2 p}{3}(1-2a)+2 (2 a-1) (p-1) p \alpha, \\
B&= 4 (p-1) p \alpha  b+b(1-\frac{4 p b}{3}). 
\label{eq:isotropic}
\end{align}
Substituting $p := p^-$ from Eq. (\ref{eq:singudepol}) in Eq. (\ref{eq:isotropic}), we find that the state obtained is the maximally mixed state, meaning that all initial states (irrespective of the parameters $a$ and $b$) become indistinguishable at this point momentarily, and the map is non-invertible at that point. Specifically, at $p^{-}$ from Eq. (\ref{eq:den}) we find that all $\nu_j$ vanish; noting that $\mathcal{E}\sigma_j = \nu_j \sigma_j$, it can be seen that vanishing $\nu_j$ corresponds to the non-invertibility of $\mathcal{E}$. 

But this singularity is not pathological in the sense that the dynamics itself remains regular, even through the singular point. In particular, the map is differentiable, in that $\dot{\mathcal{E}}\sigma_j = \dot{\nu}_j\sigma_j$ and from Eq. (\ref{eq:num}) we find that $\nu_j$ is differentiable throughout the interval $p \in [0,\frac{3}{4}]$. Moreover, this is consistent with the fact that for $p \ne p^-$, the states Eq. (\ref{eq:stateisotropic}) are distinguishable, and thus the map is invertible.

\section{Quantifying non-Markovianity: the isotropic case \label{sec:nmDepol}}

The basic dependence of the degree of non-Markovianity of the above channels on the parameters $l, m$ and $n$ can be studied more easily by considering the isotropic case, determined by the single non-Markovian parameter $\alpha$. In particular, we shall study the Hall-Cresser-Li-Andersson (HCLA) measure \cite{hall2014canonical} and the so-called SSS measure, recently introduced in Ref. \cite{shrikant2020temporal}.

In this case, all the decay rates $\gamma_1(p)$, $\gamma_2(p)$ and $\gamma_3(p)$ are equal and given by:
{\small	\begin{align}
\gamma(p) 
= \frac{\alpha  (3-6 p)+1}{(p^{-}-p)(p^{+}-p)}.
\label{eq:depolgamma}
\end{align}}
In view of Eq. (\ref{eq:asymgammas}), Eq. (\ref{eq:singuasym}) correspondingly reduces to
\begin{align}
p^{-} &\equiv \frac{-\sqrt{9 \alpha ^2-3 \alpha +1}+3 \alpha +1}{6 \alpha }.
\label{eq:singudepol}
\end{align}		
Here $p^{-}$ is the single singularity that occurs in elements of this family of channels.

The form of Eq. (\ref{eq:depolgamma}) shows that, as with the anisotropic case, the sign of the decay rate discontinuously flips from positive to negative at $p^{-}$, after which it continuously decreases in magnitude. However, whether this eventually leads to a positive rate depends on the specific value of $\alpha$, as discussed below. In particular, note that in Eq. (\ref{eq:depolgamma}), the denominator remains negative for $p> p^{-}$. On the other hand, the numerator monotonically falls with $p$. If $\alpha$ is large enough to make the numerator negative, then for such sufficiently large values, the asymptotic rate $\gamma$ becomes positive. This is determined as follows.  The point at which the numerator vanishes is
\begin{align}
\tilde{p} &= \frac{1 + 3 \alpha}{6 \alpha}.
\label{eq:phash}
\end{align}
Thus, the range of $\alpha$ for which $\tilde{p} < \frac{3}{4}$ would correspond to the failure of the QENM condition.  Observe that $\frac{d\tilde{p}}{d\alpha} = -\frac{1}{6\alpha^2} < 0$, and $\tilde{p} = \frac{3}{4}$ corresponds to $\alpha=\frac{2}{3}$. Therefore, the range $\alpha \in [\frac{2}{3},1]$ corresponds to the failure of the QENM condition.
In Figure \ref{fig:ratio_singu}, $\tilde{p}$ would correspond to the point where the rising (i.e., second) curve meets the $y=0$ line. If $\tilde{p} < \frac{3}{4}$, then the channel decay rate becomes positive asymptotically.


Recently, Ref. \cite{shrikant2020temporal} introduced a measure (which may be called the SSS measure for convenience) of non-Markovianity based on the deviation of the generator of a map giving rise to a quantum dynamical semigroup (QDS) structure. This measure expresses the idea that QDS form is equivalent to the notion of memorylessness wherein the \textit{form} of the dynamical map remains invariant in time over the evolution, and captures a notion of Markovianity stronger than CP-divisibility. The intermediate infinitesimal map is given by $(\delta\mathcal{E})\rho(p) = \mathcal{T}\exp\left(\int_p^{p+dp}\mathcal{L}(s)ds\right)\rho(p)=  (1+\mathcal{L}(p)dp)\rho(p). $ Let $\mathcal{E}^\ast(p) \equiv e^{p\mathcal{L}^\ast}$ represent the map representing semigroup evolution. Then the distance between two intermediate infinitesimal maps is:  $\delta\mathcal{E}(p) - \delta\mathcal{E}^\ast(p) = (\mathcal{L}(p) - \mathcal{L}^\ast)dp$, where $\mathcal{L}(p)$ is the generator of the dynamics and that pertaining to the semigroup form (which is time-independent). The required weaker-than-CP-divisible measure is based on the deviation from the semigroup form:
\begin{equation}
N_{SSS} = \min_{\mathcal{L}^\ast} \frac{1}{T} \int_0^T \Vert \hat{\mathcal{L}}(p) -  \hat{\mathcal{L}^\ast}\Vert dp,
\label{eq:fqds}
\end{equation}
where $\Vert A \Vert = {\rm Tr}\sqrt{AA^{\dagger}}$ is the trace norm of a matrix $A$, and $\hat{A}$ is the Choi matrix corresponding to operator $A$. For the channel (\ref{eq:nmdepol2}),  with $l=m=n\equiv \alpha$, where $ 0 < \alpha \le 1 $, the measure (\ref{eq:fqds}) reads, 
\begin{equation}
N_{SSS} = \min_{\gamma^\ast} \frac{4}{3} \int_0^\frac{3}{4} 10 \vert \gamma(p) - \gamma^\ast\vert dp,
\label{eq:fqds2}
\end{equation}
where $\gamma(p)$ is given by the Eq. (\ref{eq:depolgamma}), and $\gamma^\ast$ is the constant decay rate obtained when $\alpha =0$, which pertains to semigroup dynamics. In practice, it may be convenient to use the family semigroup limit value of $\gamma^\ast$, which is found to be $\frac{c}{4}$, given that the parametrization of time $p$ takes the form $p(t) = \frac{3}{4}(1- e^{-ct})$, where $c$ is a real number defining the strength of decay. 

In Ref. \cite{shrikant2020temporal}, the quantification Eq. (\ref{eq:fqds}) does not explicitly consider the occurrence of singularities. Here, we point that the quantification can be straightforwardly extended to the latter case, such as in the present work, by replacing the rate $\gamma(p)$ with a suitable re-normalized value. Here we use $\gamma(p) \rightarrow \gamma^\prime(p) \equiv \frac{|\gamma(p)|}{1+|\gamma(p)|}$. 

\begin{figure}
	\centering
	\includegraphics[width=7cm]{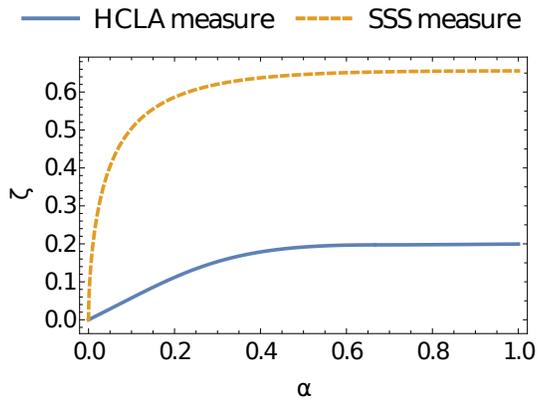}
	\caption{ Combined plot of $ N_{\rm  HCLA}$ in the Eq. (\ref{eq:intpart1} \& \ref{eq:intpart2}) and that of $N'_{SSS}$  in  the Eq. (\ref{eq:fqds2}) as a function of $\alpha$. Note that both measures increase as the non-Markovian parameter $\alpha$ increases, and never decrease. The positivity and contractive nature of the measures, for all the values of $0 < \alpha \le 1$, confirms that the map is CP.}
	\label{fig:sss_hcla}
\end{figure}

Furthermore, the resulting measure  (\ref{eq:fqds2}) itself can be suitably normalized as $N'_{SSS} = \frac{N_{SSS}}{1+N_{SSS}}$ to confine it to the range $[0,1]$. The doubly renormalized weak non-Markovianity measure against $\alpha$ is plotted as the dashed line in Fig. (\ref{fig:sss_hcla}), depicting a monotonic increase of the measure with $\alpha$. 

The renormalization technique adopted therefore serves the purpose not only to tame the singularity, but also to ensure that the non-Markovianity measure within a given family of non-Markovian channels lies between 0 (the Markovian limit in the sense of Ref. \cite{shrikant2020temporal}) and 1 (the maximally non-Markovian in the family). 

The non-Markovian behavior of this family of channels can be physically understood as follows. From Eq. (\ref{eq:nmdepol2}), we find that $\alpha$ (more generally, $l, m$ or $n$) represents the  degree by which the channel governed by a monotonic decoherence function $p(t)$ overshoots the point of maximal decoherence, and thus corresponds to a region of recurrence. This stands in contrast to the type of non-Markovianity channels we presented as elementary examples in section \ref{sec:examples}, where memory effects are incorporated easily by employing non-monotonic functions in the decoherence parameter $p(t)$. As $\alpha$ is increased, the channel's overshoot is greater, and correspondingly the greater is the non-Markovianity, as reflected in Fig. (\ref{fig:sss_hcla}). 

A point worth mentioning here is that the above measure yields a non-vanishing value even for CP-divisible channels that deviate from the semigroup structure, as discussed in Ref. \cite{shrikant2020temporal}. Furthermore, the main trend captured above, that of non-Markovianity increasing monotonically as a function of the parameter $\alpha$, is also reflected in the HCLA measure, depicted by the solid curve in Fig. (\ref{fig:sss_hcla}).

The non-Markovianity of a channel in this family according to the measure of \cite{hall2014canonical} is:
{\small		\begin{align}
N_{\rm  HCLA}(\alpha)  & :=  \int_{p : \gamma(p) < 0} \gamma'(p)dp \\
&= 	\left\{	\begin{array}{ll}
\int_{p^{-}}^{3/4}	\gamma'(p) dp \,;   
&\mbox{~for $ 0 <\alpha \le \frac{2}{3}$} \\\\
\int_{p^{-}}^{\tilde{p}}	\gamma'(p) dp \,;
&\mbox{~for $\frac{2}{3}$} < \alpha \le 1 
\end{array}\right.
\label{eq:intgamma1}
\end{align}}
where, $ \gamma^\prime(p) \equiv \frac{-\gamma(p)}{1-\gamma(p)} $ is the normalized version of $ - \gamma(p)$ in the time-local master equation. 

For example, consider the case of $ \alpha = 0.6$, for which the lone singularity is $ p^{-} \approx 0.343$, after which the decay rate is negative. Note that the rate $\gamma$ has the same behavior as the ratio function depicted in Figure \ref{fig:ratio_singu}. The non-Markovianity measure corresponding to the two expressions in Eq. (\ref{eq:intgamma1}) are, respectively		
\begin{widetext}
	\begin{subequations}
		\begin{align}
		N_{\rm HCLA}({\tiny \alpha \le \frac{2}{3}}) 	& = \frac{1}{4} \log \left(\left| \frac{\frac{3 \alpha }{4}+1}{q_1}\right| \right) +  \frac{3}{2}\left(\frac{\alpha  \left(\tan ^{-1}\left(\frac{3 \alpha -2 q_1}{q_2}\right)-\tan^{-1}\left(\frac{6 \alpha -2}{q_2}\right)\right)}{q_2}\right), \label{eq:intpart1} \\
		N_{\rm HCLA}(\alpha > \frac{2}{3}) & =\frac{1}{4} \left( \log \left| \frac{3 \alpha +\frac{1}{3 \alpha }-1}{q_1} \right| \right)+ \frac{3}{2}\left(\frac{ \alpha  \left(\tan ^{-1}\left(\frac{3 \alpha -2 q_1}{q_2}\right)-\tan ^{-1}\left(\frac{3 \alpha }{q_2}\right)\right)}{q_2} \right). \label{eq:intpart2}
		\end{align}
	\end{subequations}
	\end{widetext}
Here $q_1 = \sqrt{9 \alpha ^2-3 \alpha +1}$ and $q_2 = \sqrt{3 (4-15 \alpha ) \alpha -4}$. Note the expression is continuous at $\alpha=\frac{2}{3}$.   The corresponding plot of $ N_{\rm  HCLA}$ in the Eq. (\ref{eq:intpart1} \& \ref{eq:intpart2}) is given as the bold (blue) curve  in 
Figure (\ref{fig:sss_hcla}) and depicts, as expected, that the degree of non-Markovianity is an increasing function of $\alpha$.

\section{Volume of quasi-eternal non-Markovian Pauli channels \label{sec:enm}} 
For the family of channels considered here, the QENM property can be checked by verifying that $\gamma_j(p=\frac{3}{4}) < 0$ for any $j$. In the case of isotropically non-Markovian depolarizing channel (setting $l=m=n \equiv \alpha$), for $\alpha > 0$, as discussed earlier, the channel is CP-divisible for $0 \le p < p^-$ and CP-indivisible for $p^- < p \le \tilde{p}$, defined in Eq. (\ref{eq:phash}).  Since $\tilde{p} < \frac{3}{4}$ precisely for the set $0 < \alpha \le \frac{2}{3}$, the measure of isotropic NM depolarizing channels that is QENM is $\frac{2}{3}$.

In the general anisotropic case, it follows from Eq. (\ref{eq:asymgammas}) that the conditions for $\gamma_1, \gamma_2$ and $\gamma_3$ to be negative at $p \rightarrow \frac{3}{4}$ is
\begin{subequations}
\begin{align}
\frac{3}{4} (x_1 x_2 x_3) &< (x_2 x_3 + x_1 x_3 - x_1 x_2), \label{eq:asyma} \\
\frac{3}{4} (x_1 x_2 x_3) &< (x_2 x_3 - x_1 x_3 + x_1 x_2), \label{eq:asymb} \\
\frac{3}{4} (x_1 x_2 x_3) &< (-x_2 x_3 + x_1 x_3 + x_1 x_2), \label{eq:asymbc}
\end{align}
\label{eq:aymeternalconditions}
\end{subequations}
respectively, where $x_1 = (l+m)$, $x_2 = (l+n) $ and $x_3 = (m+n)$.  For a channel characterized by a specific value $l, m$ and $n$, if any one of the conditions (\ref{eq:aymeternalconditions}) is satisfied, then then channel would be QENM.  The volume (per Euclidean measure) of the set of QENM channels is that Eq. (\ref{eq:asyma}) or Eq. (\ref{eq:asymb}) or Eq. (\ref{eq:asymbc}) holds.  In the parameter space of $l,m$ and $n$, it is about 0.955. 

Also,  it would be pertinent to point out that the the channel introduced and characterized in \cite{shrikant2018non-Markovian} is quasi-eternal non-Markovian. This is reviewed in the Appendix, where it is characterized in terms of physical time $t$ instead of the formal parameter $p$. \bla

\section{Conclusions \label{sec:conclu}} This paper introduces depolarizing channels that are anisotropically non-Markovian, generalizing a previously proposed method \cite{shrikant2018non-Markovian}. These channels are characterized by up to three singularities in the generator. The three canonical decoherence rates were shown to flip sign after each singularity. Most members of the channels in the family are quasi-eternally non-Markovian (QENM), which is a broader class of non-Markovian channels than the eternal non-Markovian channels.  The  measure of QENM channels is found to be $\frac{2}{3}$ in the isotropic case, and 0.96 in the anisotropic case. This highlights a physical attribute to the isotropic and anisotropic non-Markovian depolarizing channels. 

Possible future directions would be to generalize this approach to random unitary channels, and study their singularity pattern and QENM property. Importantly, physical systems that can realize this, or even the simpler non-Markovian dephasing channels, would be of practical interest. Such maps show an interesting feature of level crossing eigenvalues, about which we discussed in depth in Ref. \cite{shrikant2018non-Markovian}. This may offer a clue about the kind of systems that may demonstrate such non-Markovian behavior in Nature. Reservoir engineering has recently witnessed good advances in various platforms for practical quantum information processing, and this provides a potential avenue to explore in this context.

\acknowledgements

SU and VNR thank Admar Mutt Education Foundation for the scholarship. RS and SB acknowledge, respectively, the support from Interdisciplinary Cyber Physical Systems (ICPS) program of the Department of Science and Technology (DST), India, Grants No.: DST/ICPS/QuEST/Theme-1/2019/14 and DST/ICPS/QuEST/Theme-1/2019/6. RS also acknowledges the support of the Govt. of India DST/SERB grant  MTR/2019/001516.

\bibliography{QENM}

\section*{Appendix}
\appendix
\section{Quasi-eternal non-Markovian dephasing \label{sec:appendix}}
We consider a non-Markovian analog of dephasing presented and analyzed in \cite{shrikant2018non-Markovian}, whose Kraus representation of map reads
\begin{align}
K_I(p)  &= \sqrt{[1  - \alpha  p](1-p)} \;  I \equiv \sqrt{[1-\kappa(p)]} I, \nonumber \\ 
K_Z(p)  &= \sqrt{[1 + \alpha (1-p)]p} \; \sigma_z \equiv \sqrt{\kappa(p)} \sigma_z,
\label{eq:nmdephase}
\end{align}
where $\sigma_z$ is Pauli Z operator. $\alpha$ is a real parameter that defines the degree of non-Markovianity of the channel and ranges from 0 to 1, and $p$ can be thought of as parametric time whose functional form may be given by $p(t) = \frac{1 - e^{-ct}}{2}$ such that when $t=0$, $p(t) = 0$ and when $t \rightarrow \infty$, $p(t) = \frac{1}{2}$. The decay rate can be read out from the canonical master equation $\dot{\rho} = \gamma(p)(\sigma_z \rho \sigma_z - \rho)$ or simply from the effect of map (\ref{eq:nmdephase}) on a qubit density matrix as $\begin{pmatrix}  \rho_{11} &  \rho_{12} Q(p)  \\  \rho_{21}Q(p)  & \rho_{22}  \end{pmatrix}$, with $Q(p) = 1 - 2\kappa(p)={\rm  exp}\{-2 \int_{0}^{p} \gamma(s) ds\}$, for all $0 \le s \le p$. Therefore, for NMD of the form (\ref{eq:nmdephase}) we find the decay rate to be
\begin{align}
\gamma(p) = -\frac{1}{2 Q(p)}\frac{dQ(p)}{dp} = \frac{1+ \alpha - 2 \alpha p}{1-2 p [1+ \alpha (1- p)]}.
\label{eq:nmrate}
\end{align}
\begin{figure}
	\centering
\includegraphics[width=7cm]{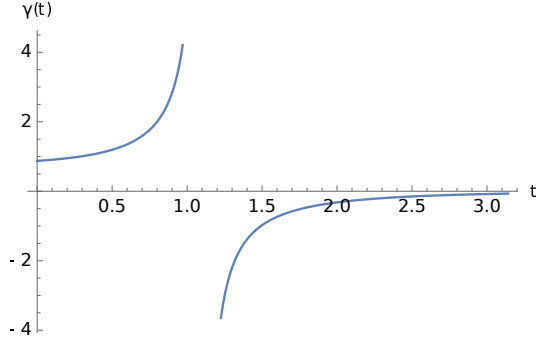}
\caption{Plot of decay rate (\ref{eq:nmrate2}) with respect physical time $t$, for non-Markovianity strength $\alpha = 0.75$ and $c=1$. In this particular case, the singularity $t^-= {\rm cosech^{-1}\alpha}$ appears at the time $t^- \approx 1.098$ units, after which the channel always remains non-Markovian though with decreasing negativity of decay rate.}
\label{fig:nmrateplot}
\end{figure}
In terms of actual physical time $t$, the decay rate (\ref{eq:nmrate}) becomes:
\begin{equation}
\gamma(t) = \frac{c (1+\alpha  e^{-ct})}{2-2 \alpha  \sinh (ct)}.
\label{eq:nmrate2}
\end{equation}  
This is plotted against time in Fig. (\ref{fig:nmrateplot}). It is interesting to note that when $t \rightarrow \infty$, then $\gamma(t) \rightarrow 0^{-}$ but never becomes zero. Hence, one may call it `quasi-eternally non-Markovian dephasing' for the reason that the channel makes transition to being non-Markovian only after the critical transition time $t^-$ and remains so ever after, see Fig. (\ref{fig:nmrateplot}). This critical time corresponds to a singularity in the time-local generator. Interestingly, the appearance of singularity in the generator is not trivial and may prompt one to look for some interesting physical features of dynamics with decay rate of the form (\ref{eq:nmrate2}), and finding such physical systems is a question we leave open. 

The singularity in time at which the decay rate blows up is found to be $t^- = \frac{{\rm cosech} ^{-1}\alpha}{c}$.  Correspondingly, at the map level, one finds a momentary non-invertibility of the intermediate map. The pair of real numbers $\{c,\alpha\}$, with $0 \le \alpha \le 1$ defines a family of non-Markovian dephasing (NMD) channels with $c$ representing the quantum dynamical semigroup limit of the family, and $c \ge 0$. The reader is referred to \cite{shrikant2018non-Markovian} for a detailed study of this channel.

\section{Appendix: Table of notations \label{sec:appendixB}}
\begin{tabular}{ |p{3cm}||p{12cm}| }
	\hline
	\multicolumn{2}{|c|}{List of notations} \\
	\hline
	Symbol  & Meaning \\
	\hline
	$\mathcal{E}$   & Quantum dynamical map   \\
	$t$& Physical time  \\
	$p(t)$ & Parametric time, which increases monotonically with time\\
	$\nu_j$ & Eigenvalue of Pauli map corresponding to the eigenoperator given by Pauli $\sigma_j$ \\
	$K_j$ & $j^{\small \rm th}$ Kraus operator\\
	$\kappa_j$& Probability defined such that $K_j(t) \equiv \sqrt{\kappa_j(t)}\sigma_j$  \\
	$\omega$& Angular frequency  \\
	$l,m,$ and $n$ & Non-Markovian parameters for the anisotropic depolarizing channel \\
	$\alpha$ & Non-Markovian parameter for the isotropic depolarizing channel \\
	$p^{\pm}$ & The point in time when singularities occur in the generator of isotropic depolarization \\
	$\tilde{p}$ & The point in time when decay rate becomes positive in the generator of isotropic depolarization \\
	$\mathcal{L^\ast}$ & Time-\textit{in}dependent Lindblad generator corresponding to semigroup evolution\\
	$\mathcal{L}(p)$ & Time-dependent Lindblad generator  \\
	$t^-$& The point in physical time when a singularity occurs in the generator  \\
	$\gamma(p)$ or $\gamma(t)$ & Time-dependent decay rate \\
	$\gamma^\ast$ & Time-\textit{in}dependent decay rate or the rate in the semigroup limit \\
	$\Lambda(p)$ & Some real function of $p$ \\
	
	\hline
\end{tabular}

\end{document}